# SENSOR SIGNAL PROCESSING USING HIGH-LEVEL SYNTHESIS AND INTERNET OF THINGS WITH A LAYERED ARCHITECTURE

CS Reddy[1] and Krishna Anand[2]

[1]Department of Mathematics, CIT - NC, VTU University, Bangalore, India
[2]Department of Computer Engineering, Anurag University, Hyderabad, India

## ABSTRACT

*Sensor routers play a crucial role in the sector of Internet of Things applications, in which the capacity for transmission of the network signal is limited from cloud systems to sensors and its reversal process. It describes a robust recognized framework with various architected layers to process data at high level synthesis. It is designed to sense the nodes instinctually with the help of Internet of Things where the applications arise in cloud systems. In this paper embedded PEs with four layer new design framework architecture is proposed to sense the devises of IOT applications with the support of high-level synthesis DBMF (database management function) tool.*

## KEYWORDS

*Network Protocols, Wireless Network, Mobile Network, Internet of Things, Reconfigurable dynamic processor, Sensor signal processing.*

## 1. INTRODUCTION

Sensor routers play a crucial role in the sector of Internet of Things (IOT) applications, in which the capacity for transmission of the network signal is limited from cloud systems to sensors and its reversal process. Consequence to this volume of the data reduction is obligatory to combat device computing functions between sensor nodes and transmitter to exchange the sufficient data with the available networks [1-3]. Hence, low power consumption and small footprints are commanded among sensor nodes to process information. One of the replacements for microcontroller units is field programmable gate arrays to optimize the footprints size, so that it is to be observed keenly the routing with configurable logic blocks and switches of look-up tables which causes placement inefficiency [4, 7]. To fabricate Field Programmable Gate Arrays (FPGA) it is wise to use High-level synthesis which will enable global optimization and recompense the limitation of Routing and Placement [5, 6].

In this paper embedded PEs with four layer new design framework architecture is proposed to sense the devises of IOT applications with the support of high-level synthesis DBMF (database management function) tool [8, 17, 19]. It exploits the repetitive high level synthesis process. Macro blocks synthesized through high-level behavioural synthesis are registered in a database before the system level synthesis, and the information in the database is used for the optimization of resource consumption through the system level synthesis [10]. In this work authors tried to investigate the dependencies of resource consumption on the granularity of coarse grained function definitions using the extended database management function of Cyber Work Bench. The evaluation results show that small footprint was achieved especially with dynamically reconfigurable technique [9, 15].

Dynamically reconfigurable processors using high-level synthesis were proposed to improve the efficiency, whereas the inefficiency of fixed mesh pointed out in still remains [11-14]. Fixed bit widths of data paths, elementary blocks, and switch matrices aiming at mass production of the devices were one example of the inefficiency. Sensors used in IOT applications have various data interfaces, such as 8, 12, 14, or 16 bits [ 17, 19]. Predefined data path between arrays of arithmetic logic units prevents behavioural synthesis tools from the optimization of layout size and the reduction in power consumption. Therefore, optimized Arithmetic Logic Units (ALU) and flexible data paths are required to embed processors in sensor units [13, 15-17].

## 2. HIGH-LEVEL SYNTHESIS TOOLS

High-level synthesis is increasingly popular for the design of high-performance and energy-efficient heterogeneous systems, shortening time-to-market and addressing today's system complexity [18, 20, 25]. Early academic work extracted scheduling, allocation, and binding as the basic steps for high-level-synthesis [22, 24, 32]. Scheduling partitions in the algorithm to control steps that are used in the model are defined the states in the finite-state machine [21, 33].

First generation behavioural synthesis was introduced by Synopsys in 1994 as behavioural Compiler and used Verilog or VHDL as input languages.  10 years later, in 2004, there emerged a number of next generation commercial high-level synthesis products which provided synthesis of circuits specified at C level to a register transfer level specification [23, 25, 26]. It was primarily adopted in Japan and Europe in the early years. As of late 2008, there was an emerging adoption in the United States. High-level synthesis (HLS) allows designers to work at a higher level of abstraction by using a software program to specify the hardware functionality [28, 29]. Additionally, HLS is particularly interesting for designing field-programmable gate array circuits, where hardware implementations can be easily refined and replaced in the target device [27, 30-32]. Recent years have seen much activity in the HLS research community, with a plethora of HLS tool offerings, from both industry and academia.

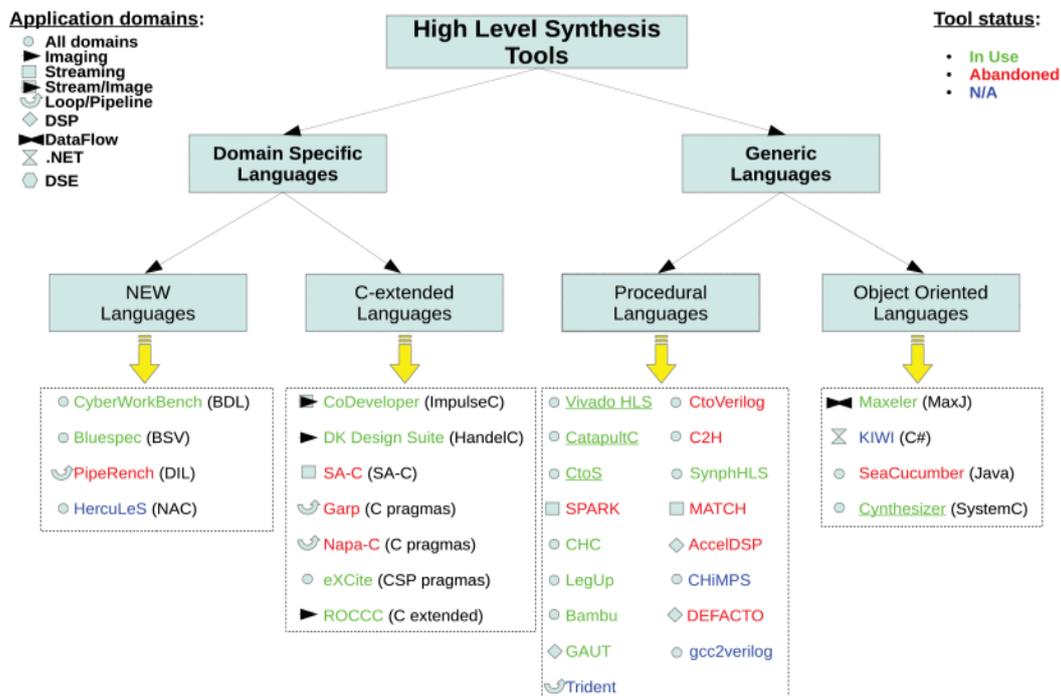

Fig.1. Classification of High-Level Synthesis Input Language.

HLS tools start from a software programmable high-level language to automatically produce a circuit specification in HDL that performs the same function. The most common source inputs for high-level synthesis are based on standard languages such as ANSI C/C++, System C and MATLAB. HLS has also been recently applied to a variety of applications (e.g., medical imaging, convolutional neural networks, and machine learning), with significant benefits in terms of performance and energy consumption.

These HLS tools can leverage dedicated optimizations or micro architectural solutions for the specific domain. However, the algorithm designer, who is usually a software engineer, has to understand how to properly update the code. This approach is usually time-consuming and error prone. For this reason, some HLS tools offer complete support for a standard HLL, such as C, giving complete freedom to the algorithm designer. High level languages mainly classified into two types that are shown in Fig. 1. One is Domain specific languages and other one is Generic languages. In Domain specific languages it consist new languages and C- extended languages, whereas, in Generic languages it consist Procedural and Object Oriented languages. HLS tool select specific language for specific applications.

## 2.1. Five Key Challenges

In this section the important crucial challenges that arise in the process of signal synthesis in the layered architecture.

- Hardware in inherently concurrent, whereas software representations are sequential. HLS must map the sequential algorithm onto concurrent hardware.
- Timing is implicit in software in the sequence of instructions used. Synchronous hardware must deal with timing constraints, with controlling and synchronizing operations at the clock cycle level.
- In software, the word length is fixed (8, 16, 32 or 64bits), but in hardware, it is usually optimized for the task being performed.
- The software model of memory is as a single block with a monolithic address space, with almost all data items stored in memory. On an FPGA, local variables are stored in registers, with multiple distributed memory blocks, each with their own independent address space. In such an environment, pointers have little meaning, and dynamic memory allocation is very difficult. Communication in software is usually through shared memory, whereas on an FPGA it relies on constructing appropriate hardware, from implicit (within stream processing), to simple token passing, to using dedicated FIFOs to manage flow control.

## 3. DATA LAYER INTERACTION ARCHITECTURE

Traditionally, the word architecture is concerned with the art or science of building, where it relates to structural concepts and to styles of design. Here the authors will use it in the sense of structured deceptions of a system from a number of compatible and complementary viewpoints which, taken together, cover functional, design, fabrication and performance issues. The term layer is used to refer to a particular descriptive viewpoint. Within a layer, descriptions will generally be hierarchic to allow the containment of complexity or the exposure of detail by suitable aggregation or decomposition techniques. Architecture(s) may be singular or plural depending on whether it will be addressed as an individual layer or the complete set of descriptions across all layers. It is clear from this definition that the architecture of a system is not just confined to some sort of functional partitioning at the front end of a development, but essentially embraces descriptions generated during all phases and stages of the development process.

The Data Interaction Architecture is a prime example of a layered architecture. It contains four layers, each of which can be regarded as a model of the system from a particular viewpoint. Fig.2. shows some small representational fragments for each of the four layers. The layers consist of I/O circuitry, fine grained layer, coarse grained function definition layer, and bypass connection layer, where they are listed from the bottom layer to the top layer, respectively. Well defined notations and technical conventions exist for each of these layers and are closely modelled on those of MASCOT, Modular Approach to Software Construction Operation and Test. Where relevant the notations include composite structures to support hierarchical representations over multiple levels within a layer. The motivation behind the layered architecture method is to provide the means of moving from a functional model of a system to an execution model.

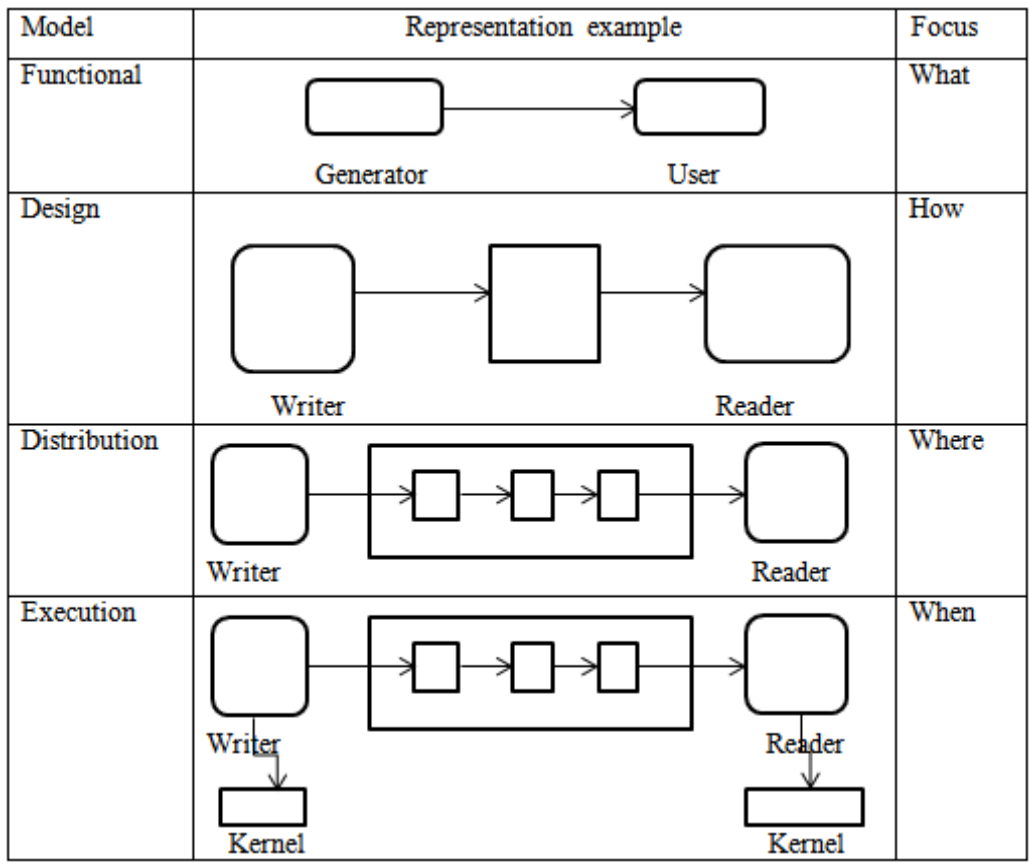

Fig.2. Data Interaction Architecture

The emphasis on the identification of well-defined interactions in the functional model, and the preservation of these speck tied interaction properties through the subsequent layers, provides the structural conformance by which traceability is achieved. It follows that the functional model is binding on subsequent development. Developers are not free to reverse engineer the functional model to allow an alternative development path in which traceability would be lost. Of course the functional model may need to be changed, either because system requirements have changed, or because detail design has shown that an alternative functional model would be better. Whatever the reason for change, consistency across the model set must be preserved.

The main distinguishing feature of Data Interaction Architecture (DIA) is the explicit representation of shared information and shared data. Here i will concentrate on the sharing implicit in the bilateral interactions between processes, leaving discussion of the more general information retention element. A summary description of the model in each layer shows how bilateral interaction properties are tracked.

## 4. ARCHITECTURE OF PROCESSING ELEMENTS

In this paper authors proposed to establish a new design framework to solve issues described in the preceding section by exploiting the binding process of high-level/behavioural synthesis.

### 4.1. Layered Architecture

The purpose of the conventional high-level synthesis tools is to generate finite state machines with data paths. Here File System Meta-Data (FSMD) referred as fine grained layer. It is expanded with the scope of high-level synthesis to coarse grained layer to exploit the functional representations of circuitries. ALUs in coarse grained layer are defined by the functional representations of high-level programming languages. Switches are added as bypass connection layer for the scope of high-level synthesis intentionally. Implicit as-built meshes of switches put constraints on high-level synthesis. In the processes are allowed as unprejudiced bypass switches to remove the meshes. Communications between the PEs are limited between adjacent PEs, whereas it is found no limitations for sensor applications with the limited communication paths.

The following explains these layers I/O circuitry is implemented with random logic gates and mixed signal I/O circuitries. They are connected to the fine grained blocks in the above layer. The fine grained layer mainly consists of finite state machines and data paths. They are replaceable to follow the context described by high-level languages. The coarse grained function definition layer is located on the fine grained layer. Optimized ALUs for a certain application are defined in this layer. An operation primitive defined in the layer corresponds to an operation code (op code) of a conventional multipoint control unit (MCU), whereas it does not have to be standardized for over many applications. As for the topmost bypass connection layer, simple bypass switch images can be specified over coarse grained blocks. The connections between inputs and outputs of PEs are configurable with the bypass switches.

### 4.2. Design Flow

The following six steps compose the design flow of the design framework to design a PE through layered scheme.

Step 1: Describe a system in high-level language.
Step 2: Define coarse grained operations. Typical dedicated functions for sensor signals are signal compensation, feature point identification for image recognition, and image compression in addition to basic arithmetic operations. The defined coarse grained operations are exploited in high-level synthesis and behavioural synthesis process.
Step 3: Generate source codes written in hardware description language through behavioural synthesis. The hard macros defined and implemented in the step 2 are exploited for generating circuitries by the functions of CWB.
Step 4: This step is identical to conventional logic synthesis.
Step 5: This step is identical to conventional layout design.
Step 6: The verification and validation step include back annotation based on the result of delay analysis.

In the work authors used a data base management function of CWB to exploit the definitions of operations in the coarse grained layer during the behavioural synthesis to treat a function as an operator. The operations can be implemented as hard macros by using custom ASIC design tools and/or LUTs of FPGAs. The operator definitions were exploited on the step 3 as macro blocks. Once macro blocks are registered in a database, CWB can use the macro blocks during the binding process of high-level synthesis automatically to reduce layout area size. The binding process of high-level synthesis is regarded as nondeterministic polynomial (NP) hard and heuristic approach was often employed. The design flow enables the automation of binding process instead of the heuristic approach.

### 4.3. Typical Design of Processing Element

Two types of context were identified with reference to semantics or lambda calculus of functional representations to define ALUs in coarse grained function definition layer. One is a configurable static context often mentioned as stored information in a file, and the other is a dynamic context as stored information in heap registers as shown in Fig. 3. The static contexts of an application are expressed with finite state machines and data paths implemented by n sets of hard- ware circuitry of PEs. The dynamic contexts of an application are specified as m sets of registers and instructions. The instruction sets are optimized for an application, and each optimized instruction set can be shared among some dynamic contexts.

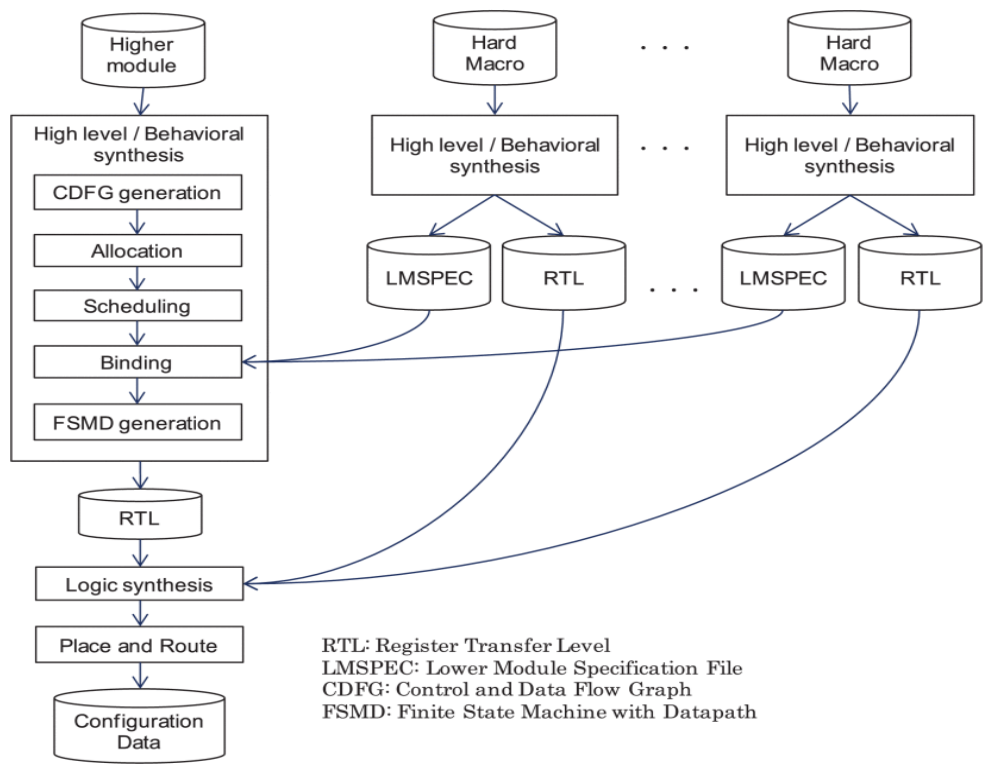

Fig.3. Modified High-Level Synthesis Process.

The program code is implemented with a specific C-language comment description /* Cyber func = operator */ for defining dynamic and static contexts, which are the sets of pairs of registers and optimized instruction sets. The pair of a register set and an instruction set can be shared among dynamic contexts using another specific C-language comment description as /* Cyber share name = NAME */. NAME is an arbitrary designation. The binding process of high-level behavioural synthesis can be controlled by these descriptions to exploit the functional representations defined in a high-level programming language.

## 5. RESULTS

Experiments are carried out to evaluate the design framework using convolution operations; those are often used for sensor applications, with following three conditions. The matrix functions of the filters are similar and the difference is the parameters of 3×3 matrices. It is observed one can evaluate the layout area size reduction results by using an FPGA, Xilinx XC7A200T FPGA, although the design framework was established aiming at improving ASIC design at first.

- ➢ Defining a convolution function as one operator.
- ➢ Implementing ALUs with basic operations, i.e., plus, minus, multiply, divide, and comparison operations, using dynamically reconfiguration technique designated as flexible reliability reconfigurable array (FRRA).
- ➢ Elaborating whole design with FPGA libraries without operator definitions and a function definition database.

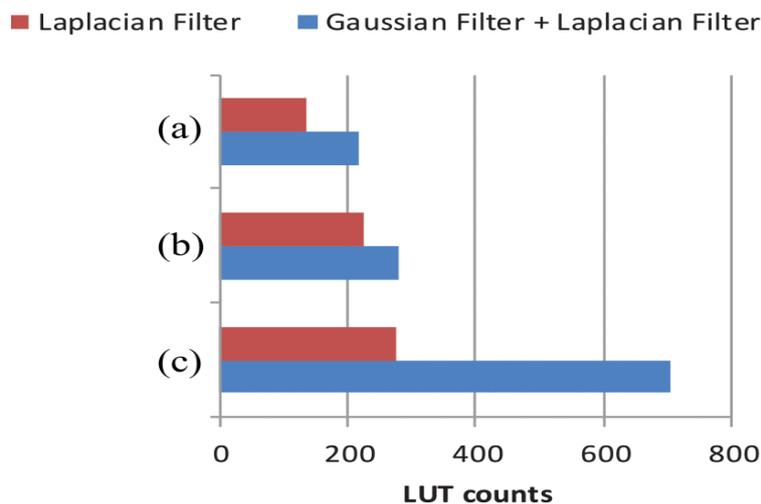

Fig.4. Comparison of LUT utilizations According to Operator Definitions

Derived LUT counts for each evaluation case are shown in Fig.4. The LUT counts of basic operations were excluded for comparison. The larger the granularity of an operation was, the less counts of LUTs were consumed. Remarkable difference of LUT counts was shown for cascaded operations. The LUT counts of cascaded operations using FPGA library was more than double compared to single operation. The results show that exploiting granularity in behavioural synthesis was carried out by CWB automatically without specifying the reuse of macro blocks explicitly. The difference between one operation and cascaded operations using FRRAs with dynamically reconfiguration technique was 22% less than the difference using the operator definition on convolution operations.

## 6. CONCLUSIONS

A novel model is developed to design a framework to reduce the footprints of programmable functions of sensing devices for IOT applications. The design framework consists of four layered structure of PE architecture and the extended database management function of a high-level synthesis tool to exploit functional representations in high-level programming languages.

The layout size reduction is useful to embed a PE into a sensing device and to provide device computing capabilities with a sensing device. The experimental results report the power consumption reduction by adopting the design framework on a Nano Bridge FPGA.